\documentclass[aps,prl,twocolumn,superscriptaddress]{revtex4} 
\usepackage[T1]{fontenc}
\usepackage{amssymb}

\usepackage{hyperref}
\hypersetup{
    bookmarks=true,         
    unicode=true,          
    pdftoolbar=false,        
    pdfmenubar=true,        
    pdffitwindow=true,      
    pdftitle={My title},    
    pdfauthor={Author},     
    pdfsubject={Subject},   
    pdfnewwindow=true,      
    pdfkeywords={keywords}, 
    colorlinks=true,       
    linkcolor=red,          
    citecolor=black,        
    filecolor=magenta,      
    urlcolor=cyan           
}

\usepackage[all]{hypcap}
\usepackage{graphicx, subfigure}
\usepackage{amsmath}
\usepackage{natbib}
\usepackage{layouts}
\usepackage{threeparttable}
\usepackage{footnote}
\usepackage{booktabs}
\usepackage{scrextend}
\usepackage{hyperref}

\begin{document}

\title{Configurational order-disorder induced metal-nonmetal transition in B$_{13}$C$_{2}$ studied with first-principles superatom-special quasirandom structure method}

\author{A. Ektarawong} 
\email{anekt@ifm.liu.se}
\affiliation{Thin Film Physics Division, Department of Physics, Chemistry and Biology (IFM), Link\"oping University, SE-581 83  Link\"oping, Sweden}
\author{S. I. Simak}
\affiliation{Theoretical Physics Division, Department of Physics, Chemistry and Biology (IFM), Link\"oping University, SE-581 83  Link\"oping, Sweden}
\author{L. Hultman}
\affiliation{Thin Film Physics Division, Department of Physics, Chemistry and Biology (IFM), Link\"oping University, SE-581 83 Link\"oping, Sweden}
\author{J. Birch}
\affiliation{Thin Film Physics Division, Department of Physics, Chemistry and Biology (IFM), Link\"oping University, SE-581 83 Link\"oping, Sweden}
\author{B. Alling}
\affiliation{Thin Film Physics Division, Department of Physics, Chemistry and Biology (IFM), Link\"oping University, SE-581 83  Link\"oping, Sweden}
\affiliation{Max-Planck-Institut f\"{u}r Eisenforschung GmbH, D-40237 D\"{u}sseldorf, Germany}


\begin{abstract} 
\indent{Due to a large discrepancy between theory and experiment, the electronic character of crystalline boron carbide B$_{13}$C$_{2}$ has been a controversial topic in the field of icosahedral boron-rich solids. We demonstrate that this discrepancy is removed when configurational disorder is accurately considered in the theoretical calculations. We find that while ordered ground state B$_{13}$C$_{2}$ is metallic, configurationally disordered B$_{13}$C$_{2}$, modeled with a superatom-special quasirandom structure method, goes through a metal to non-metal transition as the degree of disorder is increased with increasing temperature. Specifically, one of the chain-end carbon atoms in the CBC chains substitutes a neighboring equatorial boron atom in a B$_{12}$ icosahedron bonded to it, giving rise to a B$_{11}$C$^{e}$(BBC) unit. The atomic configuration of the substitutionally disordered B$_{13}$C$_{2}$ thus tends to be dominated by a mixture between B$_{12}$(CBC) and B$_{11}$C$^{e}$(BBC). Due to splitting of valence states in B$_{11}$C$^{e}$(BBC), the electron deficiency in B$_{12}$(CBC) is gradually compensated.}

\end{abstract}

\maketitle 

\section{I. Introduction}
\indent{Boron carbide, a covalent solid, is a promising candidate for a wide range of applications, e.g. lightweight armor plates, cutting tools, thermoelectric conversion and a neutron absorbing material in nuclear reactors~\cite{thevenot1990}. Boron carbide has also been realized in a neutron detector application due to a high cross section for thermal neutron reaction of $^{10}$B~\cite{Emin2005, Caruso2006, Lacy2011, Carina2012}. Experimental findings~\cite{Clark1943, Yakel1975, Morosin1986, Morosin1995} reveal a fundamental crystalline structure of boron carbide with 12-atom icosahedra placed at vertices of a rhombohedral unitcell (\emph{R$\bar{3}$m} space group) coexisting with a 3-atom intericosahedral chain aligning itself along the longest diagonal in the rhombohedron. Except from a small shift in lattice spacing, the crystalline structure of boron carbide is unchanged over a wide range of compositions ($\thicksim$8-20 at.\% C), corresponding to a single-phase region~\cite{thevenot1990, Gosset1991}. Thus, a configurationally disordered solid solution of boron and carbon atoms within the structural units are inevitable to take place. Consequently, to get an understanding of how such a substitutional disorder influences the properties of boron carbide, knowing how boron and carbon atoms are distributed to form boron carbide is of importance for its technological applications. Unfortunately, experimentally identifying the exact atomic positions of carbon atoms in these compounds, e.g. by x-ray diffraction technique, is extremely difficult due to the fact that the atomic form factors of boron and carbon are very similar. This has become a controversial issue about the structural changes of boron carbide as the carbon concentration changes, in particular for B$_{13}$C$_{2}$ with $\thicksim$13.33 at.\% C~\cite{Domnich2011}.}\\
\indent{It is generally accepted that at the carbon-rich limit, i.e. B$_{4}$C with 20 at.\% C, the structural units of boron carbide are dominated by B$_{11}$C$^{p}$(CBC), where \emph{p} denotes the polar sites of icosahedra~\cite{Bylander1990, Vast1999, Mauri2001}. This presumption is also in line with results from our recent study of configurational disorder in B$_{4}$C~\cite{Annop2014}. As the carbon concentration decreases, approaching B$_{13}$C$_{2}$, the replacement of carbon by boron can take place either within the icosahedra or in the chains, resulting in two different models: B$_{12}$(CBC) or B$_{11}$C(BBC), respectively, for B$_{13}$C$_{2}$. The former model is evident mainly by \emph{ab initio} calculations~\cite{Bylander1991, Saal2007, Vast2009}, where it is shown to have considerably lower energy as compared to the B$_{11}$C(BBC)~\cite{Bylander1991, Shirai2014}. Meanwhile the latter model is consistent with the analyses of structural data from x-ray diffraction~\cite{Aselage1992} and Raman spectra~\cite{Tallant1989, Aselage1997, Aselage1998} of boron carbide at different at.\% C in its single-phase region.}\\
\indent{Moreover, the calculations predict that the presumed B$_{12}$(CBC) is metallic~\cite{Arm1983, Bylander1991, Vast2004} due to its electron deficiency (1 hole/unitcell). The experimental observations~\cite{Wood1984, Zuppiroli1991} however reveal that, throughout the single-phase region, boron carbide is a semiconductor, thus giving rise to a discrepancy between theory and experiment. According to a study of the bonding nature of B$_{13}$C$_{2}$ by Balakrishnarajan \emph{et al.}~\cite{Musiri2007}, they suggested that the semiconducting behavior in B$_{13}$C$_{2}$ originates from unavoidable boron/carbon substitutional disorder, leading to localization of electronic states. This seems corresponding to the statement proposed by Schmechel~\cite{Werheit2000} and Werheit~\cite{Werheit2007} that the electron deficiency in B$_{13}$C$_{2}$ can be compensated by defects, e.g., substitutional defects and vacancies, splitting off some valence states into a band-gap. However, no detailed suggestion or calculations demonstrating how such effects should be realized were present. A recent theoretical study conducted by Shirai \emph{et al.}~\cite{Shirai2014} proposed structural models for B$_{13}$C$_{2}$, consisting of B$_{11}$C$^{p}$(CBC), B$_{12}$(CBC), and B$_{12}$(B$_{4}$) units. According to their models, the presence of the B$_{12}$(B$_{4}$) units can partially solve the band-gap problem and the hole density reduces to 0.25 hole/unitcell, indicating less metallic character. However, their models did not demonstrate the semiconducting character for B$_{13}$C$_{2}$, the effects of specific configurations between their suggested structural units were not fully considered, and no thermodynamic stability analysis including entropic effects were performed. Accordingly, further investigations not only to resolve the inconsistencies between experimental and theoretical studies of B$_{13}$C$_{2}$, especially the electronic structure, but also to understand the influence of substitutional disorder on the properties of boron carbide, are necessary.}\\
\indent{In this work, we investigate the phase stability, within a mean-field approximation for the configurational entropy, and the properties of substitutionally disordered B$_{13}$C$_{2}$ using first-principles calculations. A random alloy theory-based superatom-special quasirandom structure (SA-SQS) method~\cite{Zunger1990, Annop2014} is used to model disordered configurations of high concentrations of low-energy defects in B$_{13}$C$_{2}$, as previously applied to B$_{4}$C~\cite{Annop2014}. We demonstrate that, at elevated temperature, B$_{13}$C$_{2}$ is thermodynamically favorable to substitutionally disorder so that the B$_{12}$(CBC) units coexist with the B$_{11}$C$^{e}$(BBC) units, where \emph{e} stands for equatorial sites. We then show that the existence of the B$_{11}$C$^{e}$(BBC) units compensates the electron deficiency in the B$_{12}$(CBC) units. As the degree of substitutional disorder increases with temperature under equilibrium condition, a transition from the ordered B$_{12}$(CBC) metallic state to the disordered B$_{13}$C$_{2}$ semiconducting state is achieved.}\\
\section{II.  Computational details}
\indent{All calculations are performed within a realm of Density functional theory (DFT) as implemented in the Vienna \emph{ab initio} simulation package (VASP)~\cite{Kresse1996, Kresse19962}. For the total-energy calculations, the projector augmented wave (PAW) method~\cite{Blochl1994} with the Perdew-Becke-Ernzerhof (PBE96) generalized gradient approximation (GGA)~\cite{Perdew1996}, as the exchange-correlation functional, is used. Regarding the equilibrium volume optimization, the internal degrees of freedom and the cell shape are fully relaxed for a set of different fixed volume calculations. For the density of states calculations, we employ three different exchange-correlation functionals, i.e. GGA-PBE96, MBJ-GGA~\cite{Becke2006, Tran2009}, and HSE06~\cite{HSE03, HSE06}. Energy cutoff of 400 eV is used for the plane-wave calculations. Meanwhile for the Brillouin zone integration, Monkhorst-Pack \textbf{k}-point mesh~\cite{Monkhorst1976}  is chosen. Since various supercell sizes are used in this work, in order to obtain a good accuracy within a reasonable computational time for different kinds of calculations, different \textbf{k} point grids are used and given in TABLE~\ref{tab:1}. Visualizations of atomic configurations are obtained with the VESTA package~\cite{Vesta2011}.}\\
\begin{table}
\caption{Different \textbf{k}-point grids for the Brillouin zone integration, which are used in:\emph{(1)} Cell optimization calculations and \emph{(2)} Density of states calculations, for different supercell sizes.}
\begin{tabular}{c @{\hspace{0.4cm}} c @{\hspace{0.4cm}} c}
\hline
\hline
\multicolumn{3}{r @{\hspace{1.7cm}}}{\textbf{k}-point grids} \\
\cmidrule(r){2-3}
Cell size (\#atoms) & Cell optimizations & Density of states\\
\hline
1x1x1 (15) & 5x5x5 & 9x9x9 \\
 & & 5x5x5(HSE06)\\
2x1x1 (30) & 5x5x5 & 9x9x9 \\
2x2x2 (120) & 5x5x5 & 9x9x9 \\
4x3x3 (540) & 3x3x3 & 5x5x5 \\
4x4x3 (720) & 3x3x3 & 3x3x3 \\
\hline
\hline
\end{tabular}
\label{tab:1}
\end{table}
\section{III. Dilute substitutional defects}
\begin{figure}[h]
                \centering
                \includegraphics[width=0.5\linewidth]{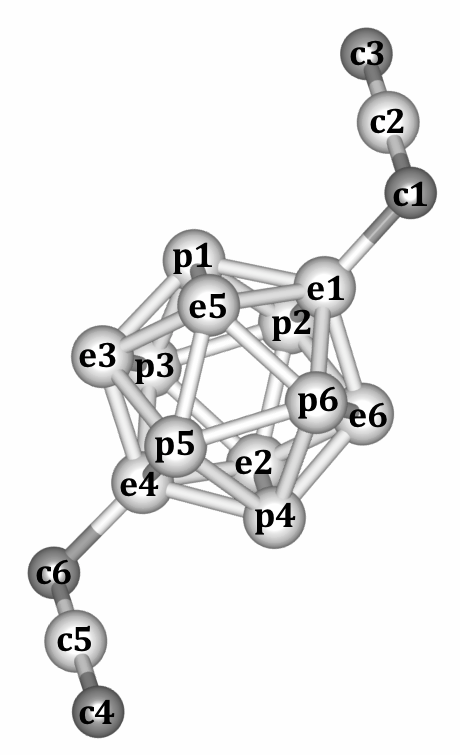}
                \caption{A 12-atom icosahedron and two 3-atom intericosahedral chains, illustrating the structural units of the predicted ground state of B$_{13}$C$_{2}$. Light and dark spheres represent boron and carbon atoms, respectively. The notations \emph{p}(1-6) and \emph{e}(1-6) stand for two different crystallographic sites in the icosahedron, i.e. polar (bonded to neighboring icosahedra) and equatorial (bonded to intericosahedral chains) sites, respectively. On the one hand, the notations \emph{c}(1-3) denote the chain sites, which are, in fact, equivalent to \emph{c}(4-6).}
                \label{fig:1}
        \end{figure}

\indent{Due to the complexity of the 15-atom structural unit and the similarity of boron and carbon atoms, numerous kinds of substitutional defects are conceivable. Taking them all into account in modelling a substitutionally disordered B$_{13}$C$_{2}$ is a difficult task. It is thus necessary to single out defects, which are low in energy and likely to appear in the structure at high concentrations. In this way, substitutionally disordered configurations of B$_{13}$C$_{2}$ can be properly modelled. We first consider different kinds of dilute substitutional defects, involved with only 2 or 4 atoms, in a 2x2x2 supercell consisting of eight B$_{12}$(CBC) units, which is believed to be the configurational ground state of B$_{13}$C$_{2}$ (see Fig.~\ref{fig:1}). We note that substitutional defects, considered in this section, are generated only by swapping boron and carbon atoms within structural units, which does not alter the global stoichiometry. Thus, all defective structures have the stoichiometry of B$_{13}$C$_{2}$. Substitutional defect formation energies, denoted by $\Delta$E$_{defect}$, are calculated by the equation;}
	\begin{equation}
		{\Delta}E_{defect}=E_{{defect}}-E_{{B_{12}(CBC)}},
		\label{eq:1}
	\end{equation}
where E$_{defect}$ and E$_{{B_{12}(CBC)}}$ are defined as the total energy of the defective structure and the ground state total energy, respectively. For each defective structure, we find that volume optimization of the supercell reduces $\Delta$E$_{defect}$ by less than 5\%. The volume is thus kept fixed at the equilibrium volume of the ground state, and during the total energy calculation, only the internal degrees of freedom are allowed to relax. $\Delta$E$_{defect}$ for different types of substitutional defects are listed in TABLE~\ref{tab:2}. We note that all considered defective structures are less energetically favorable as compared to B$_{12}$(CBC), but still, in the considered dilute concentration, stable against a phase decomposition into $\alpha$-boron and diamond.\\
\begin{table}
\caption{Substitutional defect formation energies, $\Delta$E$_{defect}$, in a 2x2x2 supercell with respect to the non-defective structure of B$_{13}$C$_{2}$, i.e. B$_{12}$(CBC). The superscripts of the notations in the parentheses correspond to the positions given in FIG.~\ref{fig:1}.}
\begin{tabular}{l @{\hspace{0.5cm}} c}
\hline
\hline
Defective structure & $\Delta$E$_{defect}$ (eV/defect)\\
\hline
Non-defective structure & \\
B$_{12}$+(C-B-C) & 0 \\
\hline
Disordered chain & \\
B$_{12}$+(B$^{(c1)}$-C$^{(c2)}$-C) & 2.570 \\
B$_{12}$+(C-C$^{(c2)}$-B$^{(c3)}$) & 2.570 \\
\hline
Polar sites & \\
B$_{11}$C$^{(p1)}$+(B$^{(c1)}$-B-C) & 1.173 \\
B$_{11}$C$^{(p1)}$+(C-B-B$^{(c3)}$) & 0.935  \\
B$_{11}$C$^{(p1)}$+(B$^{(c1)}$-C$^{(c2)}$-B$^{(c3)}$) & 3.812  \\
\hline
Equatorial sites & \\
B$_{11}$C$^{(e1)}$+(B$^{(c1)}$-B-C) & 0.605 \\
B$_{11}$C$^{(e1)}$+(C-B-B$^{(c3)}$) & 1.548 \\
B$_{11}$C$^{(e5)}$+(B$^{(c1)}$-B-C) & 1.666 \\
B$_{11}$C$^{(e1)}$+(B$^{(c1)}$-C$^{(c2)}$-B$^{(c3)}$) & 0.911  \\
B$_{11}$C$^{(e1)}$+B$_{11}$C$^{(e4)}$+(B$^{(c1)}$-B-B$^{(c3)}$) & 3.530 \\
\hline 
Bipolar defect & \\
B$_{10}$C$_{2}$$^{(p1, p4)}$+(B$^{(c1)}$-B-B$^{(c3)}$) & 1.517 \\
B$_{10}$C$_{2}$$^{(p1, p4)}$+(B$^{(c1, c6)}$-B-C)$_{2}$& 2.027 \\
B$_{10}$C$_{2}$$^{(p3, p6)}$+(B$^{(c1, c6)}$-C-C)$_{2}$ & 2.185 \\
\hline
Biequatorial defect & \\
B$_{10}$C$_{2}$$^{(e1, e4)}$+(B$^{(c1, c6)}$-C-C)$_{2}$ & 1.585 \\
\hline
\hline
\end{tabular}
\label{tab:2}
\end{table}

\indent{The atomic positions of carbon and boron atoms, in this work, are denoted by superscripts \emph{p}, \emph{e}, and \emph{c}, corresponding to their positions at polar, equatorial, and chain sites, respectively, and the numbers denote the positions according to the notations in Fig.~\ref{fig:1}. For example, B$^{(c2)}$ refers to a boron atom residing at the center position of the chain unit, i.e. the \emph{c}2 position in Fig.~\ref{fig:1}. From TABLE~\ref{tab:2}, we see that a swap of C$^{(c1)}$ by B$^{(e1)}$ yields a substitutional defect with the lowest $\Delta$E$_{defect}$ of 0.605 eV, i.e. a substitution of a chain-end carbon atom (\emph{c}1) for a neighboring equatorial boron atom (\emph{e}1) in the icosahedron, denoted by B$_{11}$C$^{e}$(BBC). This formation energy is increased by approximately a factor of 1.5, if the substitution of the other chain-end carbon atom (\emph{c}3) for a chain-center boron atom (\emph{c}2) is taken into account, i.e. B$_{11}$C$^{e}$(BCB) with $\Delta$E$_{defect}$ = 0.911 eV. Instead of a substitution of C$^{(c1)}$ for B$^{(e1)}$ as in the case of B$_{11}$C$^{e}$(BBC), $\Delta$E$_{defect}$ slightly further increases as compared to that of B$_{11}$C$^{e}$(BCB), if the substitution of a chain-end carbon atom occurs at a polar site in the neighboring icosahedron, for example at the \emph{p}1 position ($\Delta$E$_{defect}$ = 0.935 eV for a substitution of C$^{(c3)}$ for B$^{(c1)}$, yielding B$_{11}$C$^{p}$(CBB) and $\Delta$E$_{defect}$ = 1.173 eV a substitution of C$^{(c1)}$ for B$^{(p1)}$, yielding B$_{11}$C$^{p}$(BBC)). On the other hand, a substitution of C$^{(c3)}$ for B$^{(e1)}$, or B$_{11}$C$^{e}$(CBB), yields even higher $\Delta$E$_{defect}$, i.e. 1.548 eV, because of, in this case, a formation of a C-C bond between the chain unit and the icosahedron. This indicates that a direct chemical bonding between carbon atoms is unfavorable in this compound in line with the earlier finding for B$_{4}$C~\cite{Annop2014}. This is also supported by a substitution of C$^{(c3)}$ for B$^{(e5)}$ ($\Delta$E$_{defect}$ = 1.666 eV), resulting in the same kind of C-C bond as in the case of B$_{11}$C$^{e}$(CBB). Another support is a very high $\Delta$E$_{defect}$ of 2.570 eV in the case of disordered chain, i.e. B$_{12}$(BCC), where there exists an intrachain C-C bond. Since the B$_{11}$C$^{e}$(BBC) type of substitutional defect is found to have the lowest $\Delta$E$_{defect}$, with other types of substitutional defects having considerably higher $\Delta$E$_{defect}$, this type of defect is likely to dominate disordered configurations of B$_{13}$C$_{2}$.}\\
\begin{figure}[h]
                \centering
                \includegraphics[width=0.7\linewidth, angle=270]{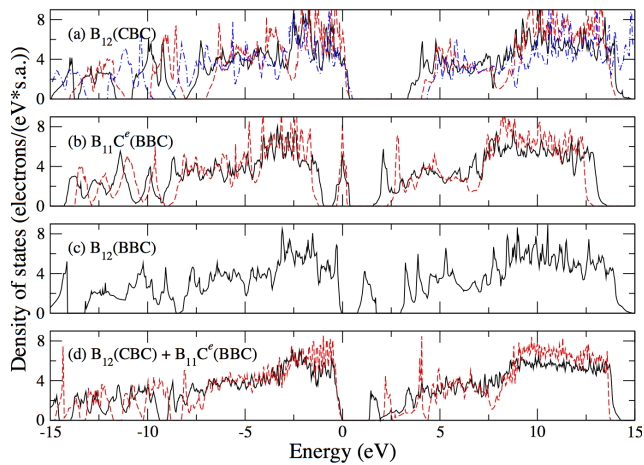}
                \caption{(Color online) Density of states of (a) B$_{12}$(CBC), (b) B$_{11}$C$^{e}$(BBC), (c) B$_{12}$(BBC), and (d) a combination of B$_{12}$(CBC) and B$_{11}$C$^{e}$(BBC). The black solid line, the red dashed line, and the blue dashed-dotted line indicate the electronic DOS, obtained by using the GGA-PBE96, the MBJ-GGA, and the HSE06, respectively, for the exchange-correlation functional. The highest occupied state is located at 0 eV for all cases.}
                \label{fig:2}
        \end{figure}
\section{IV. Splitting of valence states}
\indent{We calculate the electronic density of state (DOS) for a 15-atom unitcell of B$_{12}$(CBC), as shown in Fig.~\ref{fig:2}(a). The result, obtained from the GGA-PBE96 functional, shows the Fermi level is located in the valence band, thus resulting in a metallic behavior of the B$_{12}$(CBC) unit, due to its electron deficiency. This corresponds to the theoretical results, reported in the literature~\cite{Arm1983, Bylander1991, Vast2004}. However, experimentally, boron carbide is a semiconductor throughout the single-phase region~\cite{Wood1984, Zuppiroli1991}. We then use the MBJ-GGA and the HSE06 functionals, which are known to give a better description for electronic DOS, in particular the band-gap of semiconductors, compared to the conventional GGA-PBE96. The DOS, obtained from the latter two functionals, confirm that the B$_{12}$(CBC) unit are metallic. We thus conclude that the metallic character of the B$_{12}$(CBC) unit is, in this study, independent of the exchange-correlation functionals and the discrepancy in electronic properties between experiment and theory does not arise from inaccuracies in the used exchange-correlation functionals.}\\
\indent{Besides, we calculate the DOS of B$_{11}$C$^{e}$(BBC) in a 15-atom unitcell, as shown in Fig.~\ref{fig:2}(b). In this case, we observe splitting of two valence states into a band-gap and the Fermi level is located in the midpoint of the splitting states. Due to this valence band splitting, we consider also spin polarization in the B$_{12}$(CBC) and the B$_{11}$C$^{e}$(BBC) units, using the HSE06 functional. The results show non-zero magnetic moment in both cases. The total energy of the spin-polarized B$_{11}$C$^{e}$(BBC) is lower by 0.3 eV/unitcell, as compared to that calculated in the case of non-spin polarized calculation. On the contrary, the non-magnetic solution gives the lowest total energy to the B$_{12}$(CBC) unit. These results indicate that it might be necessary to take into account the effect of spin polarization, when calculating boron carbide configurations where the Fermi level falls into narrow defect-like states.}\\
\indent{Unlike the B$_{11}$C$^{e}$(BBC) unit, the DOS of the B$_{11}$C$^{e}$(BCB) unit (not shown) shows no splitting of valence states and the Fermi level lies within the valence band as in the case of the B$_{12}$(CBC) unit. Similarly to the B$_{12}$(CBC) unit, the B$_{11}$C$^{e}$(BBC) and the B$_{11}$C$^{e}$(BCB) units are both metallic. We suggest that such a valence band splitting in the B$_{11}$C$^{e}$(BBC) unit originates from the presence of the BBC chain. Our suggestion is supported by the valence band splitting in the boron-rich B$_{12}$(BBC) unit, as shown by Fig.~\ref{fig:2}(c). It is noted that the B$_{12}$(BBC) unit is a semiconductor, since the B$_{12}$(BBC) unit has one less electron, compared to the B$_{11}$C$^{e}$(BBC) unit, thus resulting in the two empty splitting valence states in the B$_{12}$(BBC) unit. The same kind of valence band splitting is also observed in the B$_{11}$C$^{p}$(CBB), the B$_{11}$C$^{p}$(BBC), and the B$_{11}$C$^{e}$(CBB) units (not shown).}\\
\begin{figure*}[ht!]
     \begin{center}

        \subfigure[]{%
            \label{fig:3_1}
            \includegraphics[width=0.27\textwidth]{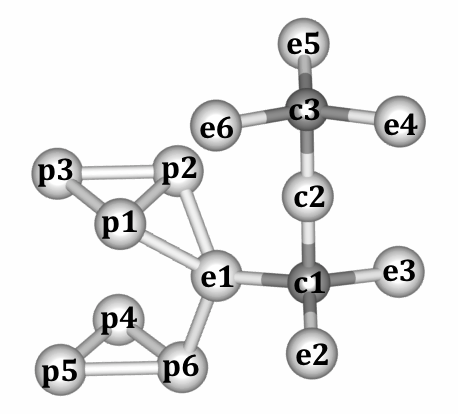}
        }%
        \subfigure[]{%
           \label{fig:3_2}
           \includegraphics[width=0.27\textwidth]{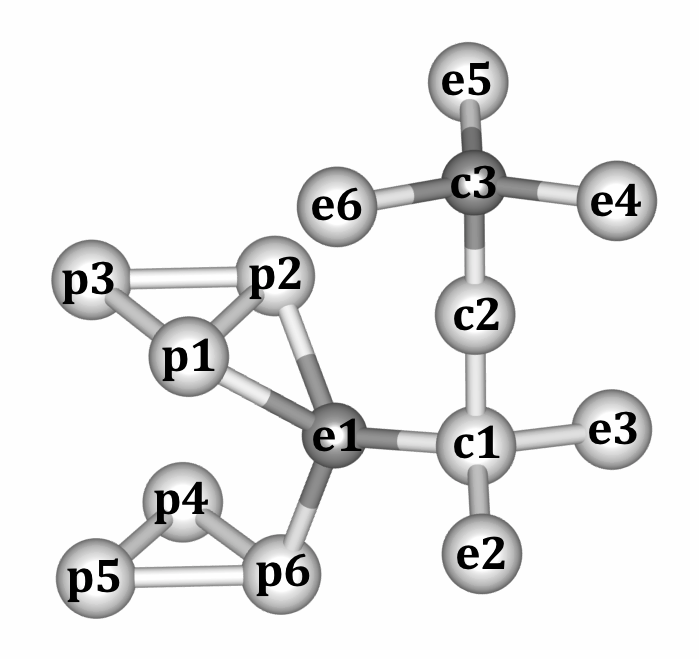}
        }
        \subfigure[]{%
           \label{fig:3_3}
           \includegraphics[width=0.27\textwidth]{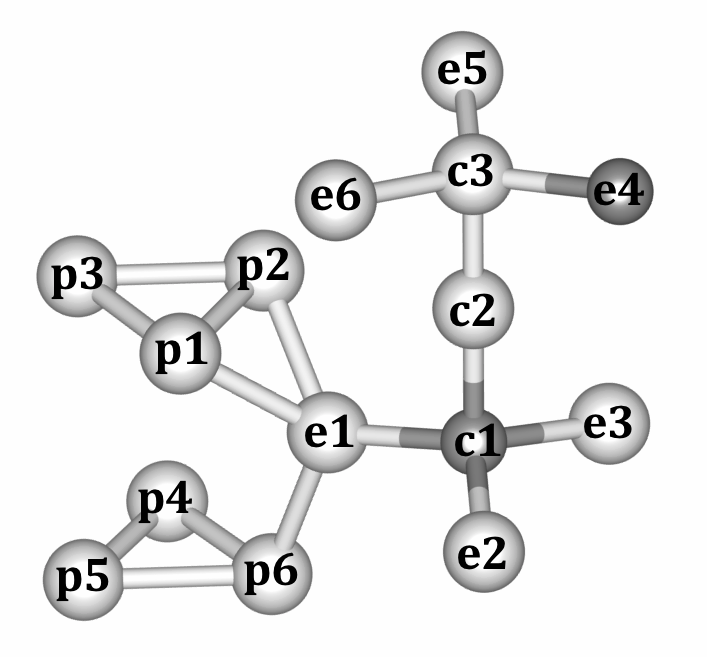}
        }\\ 
    \end{center}
    \caption{%
        A superatom basis for modelling configurationally disordered B$_{13}$C$_{2}$. Notations and colors indicate, respectively, the same crystallographic sites and the same kinds of atoms as described in Fig.~\ref{fig:1}. The types of superatoms depend on  the positions of the two carbon atoms residing in that superatom. A superatom (a) stands for the B$_{12}$(CBC) unit, while the B$_{11}$C$^{e}$(BBC) unit can be represented, for example, by a superatom (b) or (c).
     }%
   \label{fig:3}
\end{figure*}
\indent{We further investigate the electronic properties of a combination between two different units, i.e. B$_{12}$(CBC) and B$_{11}$C$^{e}$(BBC), in a 2x1x1 supercell (30 atoms). As illustrated by Fig.~\ref{fig:2}(d), the supercell of B$_{13}$C$_{2}$ exhibits a semiconducting character with a band-gap of 1.36 (2.09) eV according to a use of the GGA-PBE96 (the MBJ-GGA) functional. Also the spin polarized calculation, using the HSE06 functional reveals that it is non-magnetic. This is because an electron in the splitting states of the B$_{11}$C$^{e}$(BBC) unit occupies an empty state in the B$_{12}$(CBC) unit, thus compensating the electron deficiency. This observation is in line with the speculations by Balakrishnarajan~\cite{Musiri2007}, Schmechel~\cite{Werheit2000}, and Werheit~\cite{Werheit2007} that boron/carbon substitutional disorder could lead to localization of electronic states, compensating electron deficiency in B$_{13}$C$_{2}$. Such a compensation of electron deficiency can also take place in a combination of B$_{12}$(CBC) and B$_{11}$C$^{p}$(CBB) (B$_{11}$C$^{p}$(BBC)) with a band-gap of 1.0 (1.8) eV (not shown). Since neither the B$_{12}$(CBC) unit nor its combination with the B$_{11}$C$^{e}$(BBC) unit is magnetic, the results presented in the rest of this work will be based only on the non-spin polarized calculations.}\\
\section{V. Modeling configurationally disordered B$_{13}$C$_{2}$}
\indent{We demonstrated in section III that the formation energies of the B$_{11}$C$^{e}$(BCB), B$_{11}$C$^{p}$(CBB), B$_{11}$C$^{p}$(BBC), and B$_{11}$C$^{e}$(CBB) units in a matrix of B$_{12}$(CBC) are somewhat larger than that of the B$_{11}$C$^{e}$(BBC) unit (see TABLE~\ref{tab:2}).  In section IV, we also showed that the electron deficiency in B$_{13}$C$_{2}$ can be compensated by a mixture of B$_{12}$(CBC) and B$_{11}$C$^{e}$(BBC) with a ratio of 1:1, keeping the B$_{13}$C$_{2}$ stoichiometry. Consequently, substitutionally disordered configurations of B$_{13}$C$_{2}$ are modelled here by taking these findings into consideration. The superatom-special quasirandom structure (SA-SQS) approach~\cite{Annop2014} is used to construct configurationally disordered B$_{13}$C$_{2}$ with high concentrations of low energy defects distributed in a manner following the SQS approach for alloy theory~\cite{Zunger1990}. Fig.~\ref{fig:3} represents a chosen superatom basis that is used to model substitutionally disordered B$_{13}$C$_{2}$, mainly based on a combination of B$_{12}$(CBC) and B$_{11}$C$^{e}$(BBC). In this case, the superatom is focused on the chain, in which the equatorial sites belonging to a single icosahedron, within the basis, are replaced by the corresponding equatorial sites (\emph{e}2 to \emph{e}6 in Fig.~\ref{fig:3}) of neighboring icosahedra bonded to the original intericosahedral chain. This superatom basis allows for constructions of both the B$_{12}$(CBC) ground state unit, as well as all six displacements of the chain-end carbon atoms to their neighboring equatorial sites found to be low in $\Delta$E$_{defect}$, i.e. the B$_{11}$C$^{e}$(BBC) unit.}\\
\indent{A superatom, illustrated in Fig.~\ref{fig:3_1}, stands for the B$_{12}$(CBC) unit, meanwhile to define a superatom representing the B$_{11}$C$^{e}$(BBC) unit, we assign some constraints to the two carbon atoms within the basis in order to avoid the C-C bond between the chain unit and the icosahedron within a superatom. That is the chain-end carbon atoms can swap their positions only with the neighboring equatorial boron atoms bonded to them and only one of them is allowed to swap within the single superatom. That is, C$^{(c1)}$ can swap its position either with B$^{(e1)}$, B$^{(e2)}$, or B$^{(e3)}$ (Fig.~\ref{fig:3_2}), meanwhile C$^{(c3)}$ is allowed to swap its position with B$^{(e4)}$, B$^{(e5)}$, or B$^{(e6)}$ (Fig.~\ref{fig:3_3}). Consequently, there can be as many as seven types of superatoms considered in modelling configurationally disordered B$_{13}$C$_{2}$, consisting of the B$_{12}$(CBC) and the B$_{11}$C$^{e}$(BBC) units.}\\
\begin{table*}
\caption{Energy difference, formation energy with respect to $\alpha$-boron and diamond, supercell size, lattice constant, and electronic band-gap of different substitutionally disordered configurations of B$_{13}$C$_{2}$. In the first column, B$_{12}$(CBC), as a reference, stands for the ordered-ground state of B$_{13}$C$_{2}$. The notations \emph{e} and \emph{p} stand for the substitutionally disordered B$_{13}$C$_{2}$ with a replacement of boron atoms in the icosahedra by carbon atoms in the intericosahedral chains, respectively, at different equatorial and polar sites, indicated by the numbers, corresponding to the positions in Fig.~\ref{fig:1}, in the parenthesis. $\Delta$E, in the second column, denotes the energy difference per superatom (s.a.) with respect to the ordered state. $\Delta$E$^{form}$, in the third column, denotes the formation energy, calculated from $\alpha$-boron and diamond. The lattice parameter \emph{a} and the angle $\alpha$ are rhombohedrally averaged. E$_{g}$ denotes the electronic band-gap obtained from GGA-PBE96 (MBJ-GGA).}
\begin{tabular}{l @{\hspace{0.5cm}} c @{\hspace{0.5cm}} c @{\hspace{0.5cm}} c @{\hspace{0.5cm}} c @{\hspace{0.5cm}} c @{\hspace{0.5cm}} c @{\hspace{0.5cm}} c}
\hline
\hline
Configuration & $\Delta$E & $\Delta$E$^{form}$ & Supercell size & \emph{a} & $\alpha$ & E$_{g}$\\
& (eV/s.a.) & (eV/atom) & (Number of atoms) & (Å/unitcell) & ($^{\circ}$) & (eV)\\
\hline
B$_{12}$(CBC) & 0 & -0.092 & 1x1x1 (15) & 5.199 & 65.90 & metal (metal)\\
B$_{12}$(CBC) & 0 & -0.092 & 4x3x3 (540) & 5.197 & 65.98 & metal\\
B$_{12}$(CBC) \footnote{{Ref.~\cite{Bylander1991}} - Bylander \emph{et al.} (LDA-PP) \label{note:4}} & 0 & -0.094 & 1x1x1 (15) & 5.196 & 65.54 & metal\\
B$_{11}$C$^{e}$(BBC) & 1.778 & 0.026 & 1x1x1 (15) & 5.166 & 66.20 & metal (metal)\\
B$_{11}$C$^{e}$(BBC) \footref{note:4} & 2.093 & 0.045 & 1x1x1 (15) & 5.167 & 66.02 & -\\
\emph{e}(1) & 0.256 & -0.075 & 4x3x3 (540) & 5.169 & 65.85 & 1.04\\
\emph{e}(1, 4) & 0.248 & -0.076 & 4x3x3 (540) & 5.167 & 65.84 & 0.84\\
\emph{e}(1-3) & 0.253 & -0.076 & 4x3x3 (540) & 5.169 & 65.84 & 1.16\\
\emph{e}(1-6) & 0.258 & -0.075 & 4x3x3 (540) & 5.169 & 65.85 & 1.12 (1.65)\\
\emph{e}(1-6)+\emph{p}(1-6) & 0.420 & -0.065 & 4x4x3 (720) & 5.159 & 66.08 & 0.88\\
Exp. \footnote{Ref.~\cite{Kirfel1979} - Kirfel \emph{et al.} (XRD) \label{note:5}} & - & - & - & 5.196 & 65.62 & -\\
Exp. \footnote{Ref.~\cite{Morosin1986} - Morosin \emph{et al.} (XRD) \label{note:6}} & - & - & - & 5.185 & 65.60 & -\\
Exp. \footnote{Ref.~\cite{Werheit2006} - Werheit \emph{et al.} (Photoluminescence) \label{note:7}} & - & - & - & - & - & 2.09\\
\hline
\hline
\end{tabular}
\label{tab:3}
\end{table*}
\indent{In this study, different substitutionally disordered configurations of B$_{13}$C$_{2}$ are considered within 4x3x3 (540 atoms) and 4x4x3 (720 atoms) supercells. Results from the calculations of some selected configurations are shown in TABLE~\ref{tab:3}. The notation for each configuration, e.g., \emph{e}(1), \emph{e}(1, 4), \emph{etc.}, refers to the atomic positions within the icosahedra, corresponding to those given in Fig.~\ref{fig:1}, which are substituted by carbon atoms from the intericosahedral chains. We note that, to compensate the electron deficiency, a concentration of the B$_{12}$(CBC) unit is fixed at 50\% in all cases. Meanwhile the other 50\% is attributed to those with the BBC (or CBB) chain units, e.g., B$_{11}$C$^{e}$(BBC), B$_{11}$C$^{p}$(BBC), and B$_{11}$C$^{p}$(CBB). The latter 50\% can be equally divided, according to types of superatoms representing the units with the BBC (or CBB) chain. For example, in the \emph{e}(1-3) configuration, $\thicksim$16.67\% of equatorial boron atoms, at \emph{e}1, \emph{e}2, and \emph{e}3 positions, are equally substituted in a quasirandom manner by their neighboring carbon atoms at the \emph{c}1 position. In this case, there are three types of the B$_{11}$C$^{e}$(BBC) unit, i.e. $\thicksim$16.67\% for each. We note that in the \emph{e}(1-6)+\emph{p}(1-6) configuration, we define twelve more types of superatoms, respecting to substitution of the chain-end carbon atom (C$^{(c1)}$ or C$^{(c3)}$) for a polar boron atom within the single superatom. This is to allow not only the substitution of the chain-end carbon atoms for the equatorial boron atoms, but also for the polar boron atoms, i.e. to take into account the B$_{11}$C$^{p}$(CBB) and the B$_{11}$C$^{p}$(BBC) units, found in section III to be reasonably low in $\Delta$E$_{defect}$ and of relevance for the carbon-richer compositions approaching B$_{4}$C~\cite{Annop2014}, to model substitutionally disordered B$_{13}$C$_{2}$.}\\
\section{VI. Results and Discussion}
\subsection{a. Properties of disordered B$_{13}$C$_{2}$}
\begin{figure}[h]
                \centering
                \includegraphics[width=0.7\linewidth, angle=270]{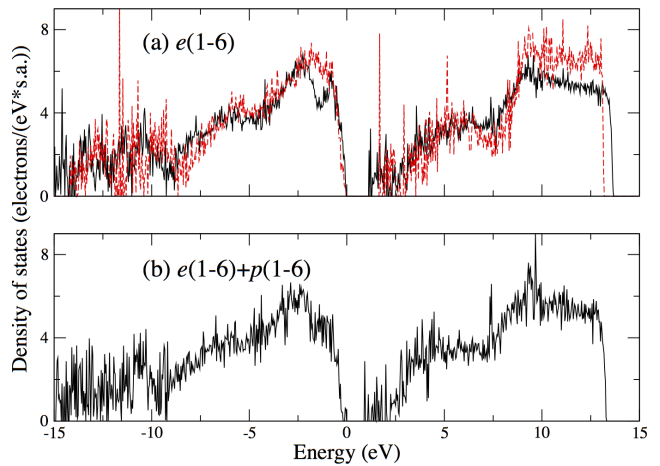}
                \caption{(Color online) Density of states of (a) the \emph{e}(1-6), and (b) the \emph{e}(1-6)+\emph{p}(1-6) configurations. The black solid line, and the red dashed line, indicate the electronic DOS, obtained by using the GGA-PBE96, and the MBJ-GGA, respectively for the exchange-correlation functional. The highest occupied state is located at 0 eV.}
                \label{fig:4}
        \end{figure}

\indent{From TABLE~\ref{tab:3}, the B$_{11}$C$^{e}$(BBC) unit has considerably higher energy as compared to the B$_{12}$(CBC) unit. This is in good agreement with the previous calculation~\cite{Bylander1991}. Except the B$_{11}$C$^{e}$(BBC) unit, the B$_{12}$(CBC) unit and all substitutionally disordered configurations of B$_{13}$C$_{2}$ are stable against a phase decomposition into $\alpha$-boron and diamond. Interestingly, all of the disordered B$_{13}$C$_{2}$ reveal a semiconducting character with a GGA-PBE96-based electronic band-gap approximately ranging between 0.8 and 1.2 eV. By taking into account the fact that the GGA-PBE96 functional regularly underestimates real band-gaps, our results are in line with the experimental band-gap of 2.09 eV, measured by photoluminescence~\cite{Werheit2006}. A better agreement of the calculated band-gap with the experiment can be achieved by a use of other exchange-correlation functionals that give a better description of electronic DOS. For instance, the band-gap of the \emph{e}(1-6) configuration becomes 1.65 eV with a use of MBJ-GGA. The electronic DOS of the \emph{e}(1-6) and \emph{e}(1-6)+\emph{p}(1-6) configurations are shown in Fig.~\ref{fig:4}. It is worth noting that the presented results are different to all previous calculations for B$_{13}$C$_{2}$, and correspond to the experimental observations that B$_{13}$C$_{2}$ is a semiconductor.}\\
        
\indent{Experimentally, boron carbide has the rhombohedral symmetry (\emph{R$\bar{3}$m}). The lattice parameters (\emph{a}, \emph{b}, \emph{c}) and also the angles ($\alpha$, $\beta$, $\gamma$) at equilibrium for all disordered configurations listed in TABLE~\ref{tab:3} are slightly deviating from each other by less than 1\%. We attribute this deviation to substitutional disorder within the supercell with a finite size. The existence of the icosahedral carbon in the B$_{11}$C$^{e}$(BBC) unit results in the monoclinic distortion as does the ground state B$_{4}$C, i.e. B$_{11}$C$^{p}$(CBC). As a result, the lattice parameters and the angles, given in  TABLE~\ref{tab:3} are rhombohedrally averaged. The averaged parameters \emph{a} and $\alpha$ for disordered B$_{13}$C$_{2}$ are slightly underestimated by less than 0.5\% and 0.3\%, respectively, when compared to the experiments~\cite{Kirfel1979,Morosin1986}. The averaged \emph{a} and $\alpha$ for the B$_{11}$C$^{e}$(BBC) unit are in agreement with the previous work~\cite{Bylander1991}, in which the averaged \emph{a} is practically identical to the disordered B$_{13}$C$_{2}$, meanwhile the averaged $\alpha$ is slightly larger with respect to the experiments~\cite{Kirfel1979,Morosin1986}. The ground state configuration, on the other hand, is more similar to the experiments for the parameter \emph{a} but its angle $\alpha$ is slightly more deviated, as compared to those of the disordered configurations. As demonstrated by Isaev \emph{et al.}~\cite{Isaev2011}, the effects of lattice vibrations, i.e. zero-point motion and finite temperature effects, can play an important role in a system of light elements, e.g. elementary boron. In such a case, the inclusion of vibrational effects corrects the underestimated equilibrium volume and gives excellent agreement with the experiment. We suggest that the underestimated lattice parameters for the disordered B$_{13}$C$_{2}$ may be originating from the neglect of lattice vibrations. Regarding the crystal symmetry, only the \emph{e}(1, 4), the \emph{e}(1-6), and the \emph{e}(1-6)+\emph{p}(1-6) configurations have, on average, the full rhombohedral symmetry with the \emph{R$\bar{3}$m} space group, while the \emph{e}(1), and \emph{e}(1-3) configurations are lacking inversion symmetry, thus yielding the \emph{R}3\emph{m} space group on average. We note that the existence of the \emph{R}3\emph{m} phase has been discussed by us in configurationally disordered B$_{4}$C, stable at elevated temperatures~\cite{Annop2014}, and more recently also by Yao \emph{et al.}~\cite{Yao2014}.}
\subsection{b. Bending and rotational effects in BBC chains}
\begin{figure}[h]
                \centering
                \includegraphics[width=0.7\linewidth]{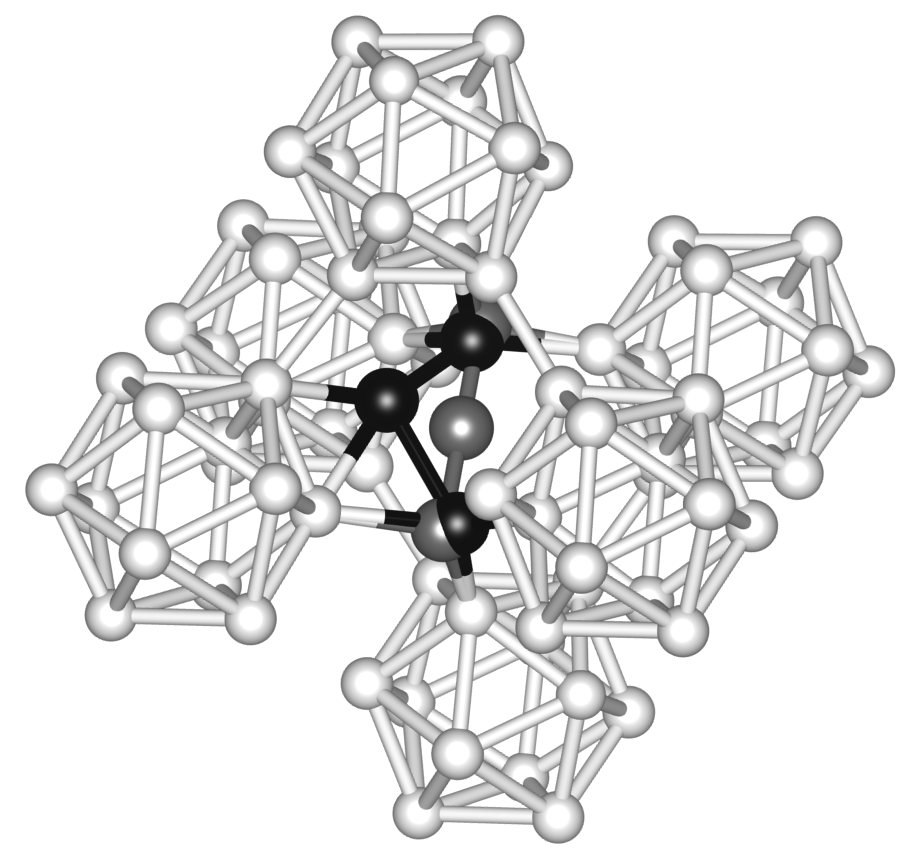}
                \caption{Illustration of a BBC chain before (grey spheres) and after (black spheres) ionic relaxations, and their neighboring icosahedra (white spheres).}
                \label{fig:5}
        \end{figure}

\indent{After the ionic relaxation, we observe that the BBC chains in the disordered B$_{13}$C$_{2}$ are bending, by which the chain-center boron atoms in the BBC chains are displaced to interstitial positions around their former positions, resulting in empty space at the chain-center positions. An illustration of the bended BBC chain is given by Fig.~\ref{fig:5}. We further investigate the bended BBC chain by 360$^{\circ}$ rotating the chain, and find that the chain-center boron atoms move toward and likely to form bonds with neighboring icosahedra bonded to the chain-end boron atoms. In addition, the chain-center boron atom is unfavorable to form bond with the icosahedral carbon atom bonded to the chain-end boron atom. These findings are in line with the work of Kwei \emph{et al.}~\cite{Morosin1996}, where they demonstrated, using neutron diffraction technique, the existence of both the vacancies at the chain-center positions and the interstitial atoms around the chain-center positions. They stated that this may originate from the presence of the rhombic B$_{4}$ or non-linear chains, such as BBB chain. However, we demonstrated in section III that the BBB chain is the high energy defect. As for the rhombic B$_{4}$ chain, it will be shown in the following section VI:e that a model with the rhombic B$_{4}$ chain, proposed by Shirai \emph{et al.}~\cite{Shirai2014} is less favorable compared to our models in the present work. We thus suggest that rather than the BBB and the B$_{4}$ chains, the vacancies at and the interstitial atoms around the chain-center positions could originate from bending of the BBC chains. Furthermore, due to the chain bending, the distance between the two chain-end atoms in the BBC chain decreases and ranges between 2.45 and 2.65 Å, which is in agreement with the value of 2.5 Å reported by Kwei \emph{et al.}~\cite{Morosin1996}. Meanwhile the distance between the two chain-end boron atoms in the bending BBB chain and the rhombic B$_{4}$ chain in our case are 1.85 Å and 2.85 Å, respectively.}\\
        
\indent{Apart from the BBC chains, we discuss about the CBC chains in the disordered B$_{13}$C$_{2}$. Unlike the bended BBC chains, the CBC chains are still linear after the relaxation. The length of the linear CBC chains in the disordered B$_{13}$C$_{2}$ is slightly different from that of the CBC chain in the B$_{12}$(CBC) unit (2.886 Å) by less than 0.25\%. Meanwhile the length of the CBC chain in the B$_{12}$(CBC) unit increases by 0.52\% approximately as compared to that of the CBC chain in the B$_{11}$C$^{p}$(CBC) unit of B$_{4}$C. This is in excellent agreement with the experimental results existing in the literature~\cite{Morosin1986, Kirfel1979, Larson1986}, in which the length of the CBC chains in B$_{13}$C$_{2}$ increases approximately from 0.21\% to 0.34\% with respect to that of the CBC chain in B$_{4}$C.}
\subsection{c. Configurational stability of the disordered B$_{13}$C$_{2}$}
\begin{figure}[h]
                \centering
                \includegraphics[width=0.7\linewidth, angle=270]{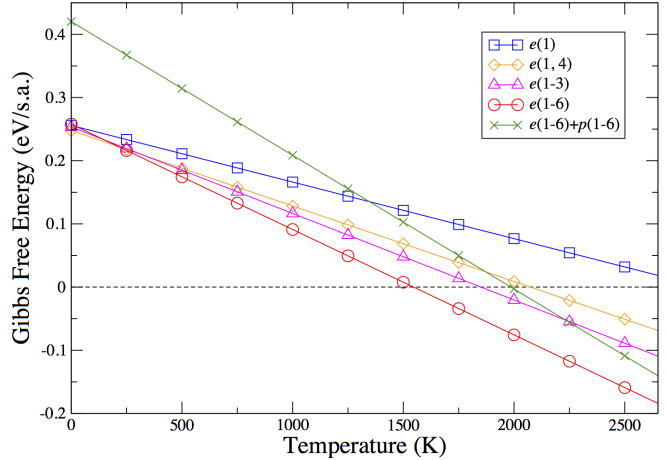}
                \caption{(Color online) Gibbs free energies for different substitutionally disordered B$_{13}$C$_{2}$ plotted as a function of absolute temperatures (at P = 0 GPa), relative to the B$_{12}$(CBC) ground state (dashed line).}
                \label{fig:6}
        \end{figure}

\indent{We also determine the stability of substitutionally disordered B$_{13}$C$_{2}$, with respect to the ordered ground state B$_{12}$(CBC). The Gibbs free energy for disordered configuration $\gamma$ at zero pressure is obtained from;
	\begin{equation}
		G^{\gamma}=E^{\gamma}-TS^{\gamma},
		\label{eq:2}
	\end{equation}
where \emph{G}$^{\gamma}$, \emph{E}$^{\gamma}$ and \emph{S}$^{\gamma}$ are referred to the Gibbs free energy, the total energy, and the configurational entropy, respectively. \emph{T} is the absolute temperature in Kelvin. Meanwhile the configurational entropy is calculated within a mean-field approximation, as given by;
	\begin{equation}
		S=-k_{B}N\sum_{i=1}^{n}x_{i}\ln(x_{i}),
		\label{eq:3}
	\end{equation}
where k$_{B}$ is the Boltzmann constant. \emph{N} and \emph{n} are defined as the number of superatom sites in the supercell and the number of superatom types included in the supercell, respectively. \emph{x$_{i}$} refers to the concentration of type-\emph{i} superatom. By investigating the rotational effects of the bended BBC chains in the disordered B$_{13}$C$_{2}$ as mentioned in section VI:b, we find that there are two ways of alignments that the bended chains can form bonds with the icosahedra and thus result in practically the same total energy. This bending-angle degeneracy provides a contribution to the configurational entropy. The extra configurational entropy from the bending and rotational degree of freedom of the BBC chains is thus included by assuming that the number of superatom types and the concentration for each superatom type, representing the B$_{11}$C$^{e}$(BBC) units in Eq.~\ref{eq:3} become twice and half, respectively. Fig.~\ref{fig:6} illustrates the Gibbs free energy, modeled with the mean-field SA-SQS approach, for different substitutionally disordered B$_{13}$C$_{2}$ plotted as a function of absolute temperature. As shown in TABLE~\ref{tab:3}, the energy differences for the \emph{e}(1), the \emph{e}(1, 4), the \emph{e}(1-3), and the \emph{e}(1-6) configurations, relative to the OS configuration, are approximately 0.25 eV/s.a., where the energy differences among them are small [on the order of 10$^{-3}$ eV/s.a.]. We find that, above 1546 K, the \emph{e}(1-6) configuration becomes stable, with respect to the ordered ground state. This is due to the higher configurational entropy for the \emph{e}(1-6) configuration, that lowers the Gibbs free energy at high temperature, more than those of the other types of disorder. The high-temperature phase of B$_{13}$C$_{2}$ can thus be represented by a mixture of the B$_{12}$(CBC) and the B$_{11}$C$^{e}$(BBC) units, in which some chain-end carbon atoms C$^{c1}$ (C$^{c3}$) swap position with one of the neighboring equatorial boron atoms B$^{e1}$, B$^{e2}$, B$^{e3}$ (B$^{e4}$, B$^{e5}$, B$^{e6}$) within the icosahedra bonded to the chain-end atoms. Even though the \emph{e}(1-6)+\emph{p}(1-6) configuration has the highest configurational entropy, indicated by its high slope in Fig.~\ref{fig:6}, among the others listed in TABLE~\ref{tab:3}, its free energy is rather high even at high temperature. This corresponds to the results obtained from the dilute substitutional defects study in section III.}\\ 
        
\begin{figure}[h]
                \centering
                \includegraphics[width=0.72\linewidth, angle=270]{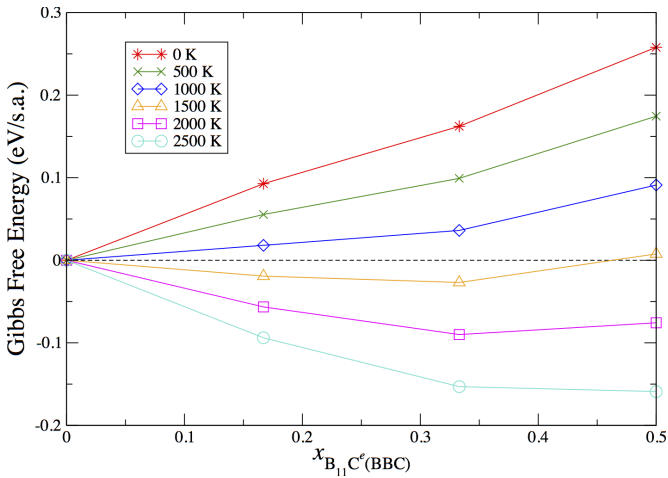}
                \caption{(Color online) Plot between the Gibbs free energies of the \emph{e}(1-6) configuration and the concentration \emph{x} of the B$_{11}$C$^{e}$(BBC) units at various absolute temperatures, relative to the B$_{12}$(CBC) ground state (dashed line).}
                \label{fig:7}
        \end{figure}

\indent{So far only the \emph{e}(1-6) configuration, consisting of 50\% of the B$_{12}$(CBC) and the B$_{11}$C$^{e}$(BBC) units, i.e. (B$_{12}$(CBC))$_{0.5}$(B$_{11}$C$^{e}$(BBC))$_{0.5}$, is considered. We thus further investigate the \emph{e}(1-6) configuration, as generally denoted by (B$_{12}$(CBC))$_{1-x}$(B$_{11}$C$^{e}$(BBC))$_{x}$, by which we vary the concentration \emph{x} of the B$_{11}$C$^{e}$(BBC) units. The concentration \emph{x}, reflecting the degree of substitutional disorder in B$_{13}$C$_{2}$, is varied between 0 (100\% of B$_{12}$(CBC)) and 1 (100\% of B$_{11}$C$^{e}$(BBC)). Fig.~\ref{fig:7} depicts a plot between the Gibbs free energy of the \emph{e}(1-6) configuration and the concentration \emph{x} at various absolute temperatures, where 0$\leq$\emph{x}$\leq$$\frac{1}{2}$. It is worth noting again that the B$_{11}$C$^{e}$(BBC) units have 6 types of degeneracy, according to the substitution of the chain-end carbon atoms at one out of the six equatorial sites, and have 12 types of degeneracy if the rotational degree of freedom of the bended BBC chains is taken into account. In this study, the concentration of B$_{11}$C$^{e}$(BBC) is sampled discretely at \emph{x} = 0, $\frac{1}{6}$, $\frac{1}{3}$, $\frac{1}{2}$, $\frac{2}{3}$, $\frac{5}{6}$, and 1. The results in Fig.~\ref{fig:7} indicate that as the temperature increases, B$_{13}$C$_{2}$ is energetically favorable to disorder, where some of the B$_{12}$(CBC) units will transform into the B$_{11}$C$^{e}$(BBC) units. According to our samplings, the \emph{e}(1-6) configuration with \emph{x} = $\frac{1}{6}$, $\frac{1}{3}$, and $\frac{1}{2}$ will be energetically favorable when the temperature is higher than 1243 K, 1350 K, and 2350 K, respectively. Also, because of such a disorder, the electron deficiency in B$_{13}$C$_{2}$ will be gradually compensated until it eventually becomes a semiconductor, as \emph{x} is approaching $\frac{1}{2}$. In our case, the hole density is reduced to 0.67, 0.33, and 0 hole/s.a. for the \emph{e}(1-6) configuration with \emph{x} = $\frac{1}{6}$, $\frac{1}{3}$, and $\frac{1}{2}$, respectively. As \emph{x} is larger than $\frac{1}{2}$, the stability of the disorderd B$_{13}$C$_{2}$ tends to drastically decrease, predicted by a relatively higher Gibbs free energy even at high temperatures. This is not only because of a constant increase of the less stable B$_{11}$C$^{e}$(BBC) units, but also a non-linear effect at \emph{x}>$\frac{1}{2}$ (not shown).}\\
        
\indent{Based on our results, a temperature as high as 2350 K is required in order to achieve a completely semiconducting B$_{13}$C$_{2}$. We note that boron carbide can be manufactured, for instance, by hot-pressing, at typically above 2000 K~\cite{thevenot1990}. This high temperature corresponds to the study of Kuzenkova \emph{et al.}~\cite{Kuzen1979} that the kinetics of recrystallization, reflecting atomic diffusion, in boron carbide starts at temperature above 2000 K due to its strong interatomic covalent bonds. Our results reveal, at 2000 K, the Gibbs free energy of the \emph{e}(1-6) configuration with \emph{x} = $\frac{1}{2}$ is higher than that with \emph{x} = $\frac{1}{3}$ only by less than 1 meV/atom, which is practically negligible. On the other hand, upon cooling, the atomic diffusion within boron-carbon compound might be limited/quenched, thus preventing it to reach the ordered metallic B$_{12}$(CBC) state. We thus suggest that this could be the main cause for B$_{13}$C$_{2}$ to behave as a semiconductor, i.e. due to the inevitable substitutional disorder between boron and carbon atoms. Our results corroborate the suggestions by Balakrishnarajan \emph{et al.}~\cite{Musiri2007}, Schmechel~\cite{Werheit2000}, and Werheit~\cite{Werheit2007}, that semiconducting behavior in B$_{13}$C$_{2}$ could arise from substitutional disorder.}
\subsection{d. Estimation of the uncertainty in transition temperature predictions}
\indent{The transition temperatures, calculated in this work, involve several approximations that can give rise to the uncertainty in predicting transition temperatures. Consequently, we estimate in this section errors for the transition temperatures, originating from different sources, and discuss how such errors affect the transition temperatures.}\\
\indent{The first source of errors originates from the approximations in DFT itself. Since the exchange-correlation energy functionals are not known exactly, calculated results, i.e. the transition temperature in our case, may either increase or decrease depending on used exchange-correlation functionals. Hautier \emph{et al.}~\cite{Hautier2012} recently modeled the computational errors within the DFT using GGA (+U for d-block metals) as the exchange-correlation functional. The errors were estimated by evaluating the reaction energies for ternary from binary oxides using DFT and compared them with experiments. The error, i.e. standard deviation, is thus approximated to be 24 meV/atom. In our case, the error can be expected to be much smaller, since the bonding character and the local atomic environment are very similar for most atoms when comparing disordered boron carbides and the ordered ones. Meanwhile in the case of Hautier \emph{et al.}, forming ternary oxides from different binary oxides does considerably change bonding characters as well as the atomic environment of most atoms. Instead, if we approximate that the DFT error in our case is around 24 meV per \emph{superatom}, the structural unit observing distinct changes, we approximate the error bar for the order-disorder transition temperature between the disordered and the ordered B$_{13}$C$_{2}$, which are respectively the \emph{e}(1-6) and the ground-state B$_{12}$(CBC), due to exchange-correlation uncertainties to be around 10\%, or 150 K.}\\
\indent{Another source of errors is attributed to the use of the SA-SQS approach. Unlike normal atoms, each superatom has an internal structure, consisting of at least 15 atoms. Consequently, the commutative property is not preserved within the SA-SQS, i.e. BA $\neq$ AB, in which A and B are superatoms of different types. To estimate the error, we consider four substitutionally disordered configurations of B$_{4}$C, in which, for each configuration, the random substitution of icosahedral carbon atoms C$^{p}$ at all six polar sites coexists with the bipolar defects (B$_{10}$C$_{2}^{p}$+B$_{12}$) at high concentrations, as proposed to be the high temperature phase of B$_{4}$C and being denoted by the notation BD+6P in our previous work~\cite{Annop2014}. The four models of BD+6P are obtained, using the same superatom basis and the same superatom types, as suggestes in Ref.~\cite{Annop2014}, but they are allowed to interchange. Due to the commutation of superatom types, each BD+6P has different internal configuration from the others, thus resulting in different total energies as well as different transition temperatures. We estimate the uncertainty by calculating a mean value of transition temperature between the BD+6P and the ground-state B$_{11}$C$^{p}$(CBC) unit and its standard deviation. The mean transition temperature and the standard deviation are 1426 K and 152.45 K, respectively. Therefore, the transition temperature of 1426$\pm$305 K is obtained with 95\% confidence intervals, i.e. the uncertainty in transition temperature caused by this approximation is $\lesssim$ 20\%.}\\
\indent{The third error is due to the use of mean-field approximation for the configurational entropy. It is known that this approximation generally overestimates the order-disorder transition temperature by 20-30\% due to neglecting short range ordering effects. An example is given by the work of Alling~\cite{Alling2014}, in which the order-disorder transition temperature of Ti$_{0.5}$Mg$_{0.5}$N was calculated both with mean-field and Monte Carlo methods based on unified cluster expansion~\cite{Alling2011}. The mean-field transition temperature was found to be 1272 K, overestimating the more accurate Monte Carlo simulation that obtained the transition temperature of 950 K by 322 K or 30\% approximately. Recently, Yao \emph{et al.}~\cite{Yao2014} studied configurational phase transition in B$_{4}$C by which they used Monte Carlo simulations. Based on their work, they proposed a pair of transitions. After the first transition at 717 K, they found that the icosahedral carbon atoms in B$_{11}$C$^{p}$ of the monoclinic ground state B$_{4}$C, represented by the B$_{11}$C$^{p}$(CBC) units, are substitutionally disordering on the three polar-up (or down) sites, resulting in the \emph{R}3\emph{m} symmetry. This is in fact equivalent to the 3PU configuration in our previous work of B$_{4}$C~\cite{Annop2014}. Compared to ours, the transition temperature in our case is slightly higher by 150 K due to the use of mean-field approximation, i.e. overestimation with 20\% approximately. For the higher transition, where icosahedral carbon atoms are disordering at all six polar sites (\emph{R$\bar{3}$m}), the difference in the transition temperature between the two works are larger, but further investigations are needed before this difference is referred only to the mean-field approximation.}\\
\indent{Perhaps the largest source of uncertainty in the present temperature prediction is the absence of explicit vibrational effects. We referred to the work of Isaev~\cite{Isaev2011} in section VI:a that the effect of lattice vibrations can play an important role in a system of light elements. Garbulsky \emph{et al.}~\cite{Garbulsky1996} demonstrated that the vibrational effect on order-disorder transition temperature can be significant, by which it tends to lower the transition temperature. Taking into account the vibrational effects is however beyond the scope of this work.}\\
\indent{Taken together, our approach can give quantitative information about the properties of disordered phases, and a qualitative description of a series of order-disorder phase transitions, but not a quantitative description of their exact transition temperatures. It does however provide the necessary starting point, in term of candidate structures, for future more elaborate investigations, in particular including vibrations. By taking into account the errors that could be arising from our approach, our transition temperatures are likely to be overestimated by several hundreds Kelvin. However, based on the discussion above, even though our transition temperatures are high, they are not going to exceed the melting temperature of the material. As a result, the disordered B$_{13}$C$_{2}$ consisting of the B$_{12}$(CBC) and the B$_{11}$C$^{e}$(BBC) units can be a representation of a high temperature phase of B$_{13}$C$_{2}$ and would be able to provide a solution to the electronic structure problem in B$_{13}$C$_{2}$.}
\begin{figure}[ht!]
     \begin{center}
        \subfigure[]{%
            \label{fig:8_1}
            \includegraphics[width=0.15\textwidth]{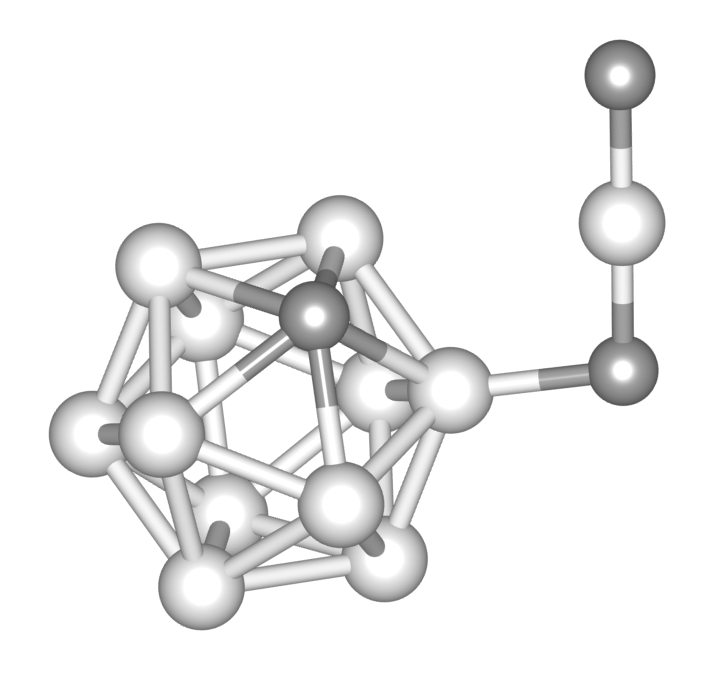}
        }%
         \subfigure[]{%
            \label{fig:8_2}
            \includegraphics[width=0.15\textwidth]{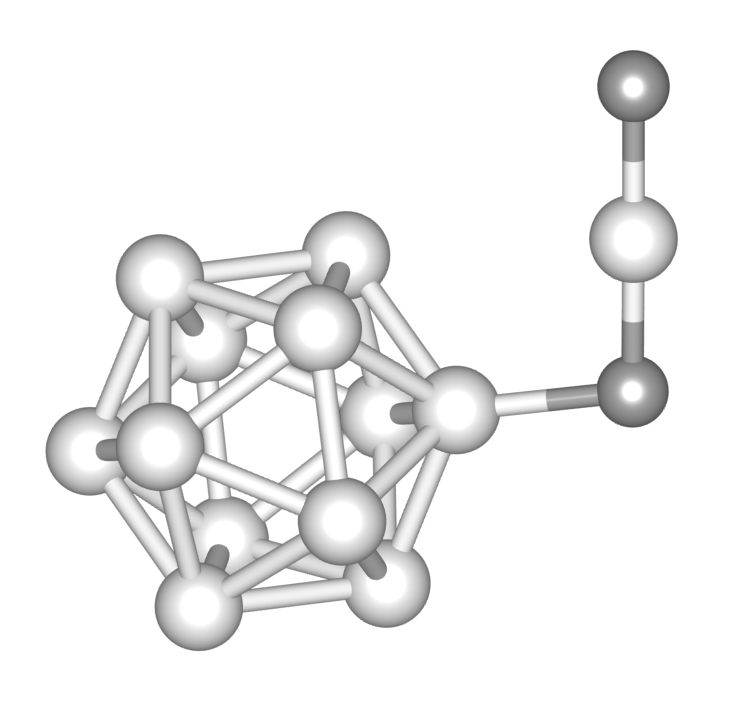}
        }%
        \subfigure[]{%
           \label{fig:8_3}
           \includegraphics[width=0.165\textwidth]{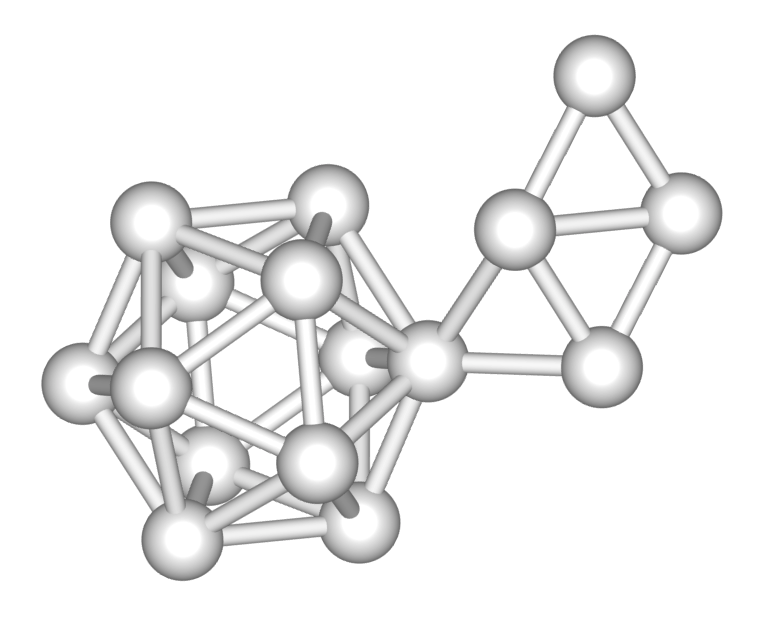}
        }\\ 
    \end{center}
    \caption{%
        Three types of superatoms, used to construct structural models for B$_{13}$C$_{2}$ proposed by Shirai \emph{et al.}~\cite{Shirai2014}: (a) B$_{11}$C$^{p}$(CBC), (b) B$_{12}$(CBC), and (c) B$_{12}$(B$_{4}$) with a 4-atom rhombic chain. Light and dark spheres represent boron and carbon atoms, respectively.
     }%
   \label{fig:8}
\end{figure}
\subsection{e. Comparison with models proposed in the literature}
\begin{figure}[h]
                \centering
                \includegraphics[width=0.72\linewidth, angle=270]{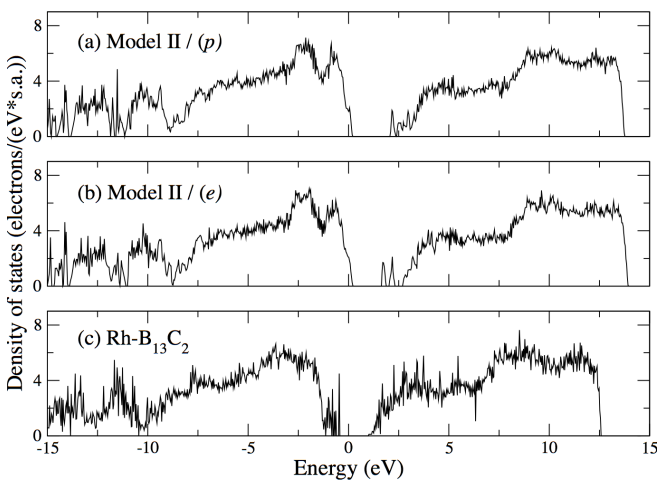}
                \caption{Density of states of (a) Model II/(\emph{p}), (b) Model II/(\emph{e}), and (c) Rh-B$_{13}$C$_{2}$, obtained by using the GGA-PBE96 for the exchange-correlation functional. The Fermi level is located at 0 eV.}
                \label{fig:9}
        \end{figure}

\indent{Recently, Shirai \emph{et al.}~\cite{Shirai2014} proposed an alternative structural model for B$_{13}$C$_{2}$, consisting of 3 different types of structural units, i.e. B$_{11}$C$^{p}$(CBC), B$_{12}$(CBC), and B$_{12}$(B$_{4}$) with a 4-atom rhombic (B$_{4}$) chain (see Fig.~\ref{fig:8}). We were thus inspired, with a use of SA-SQS, to examine the model II in Ref.~\cite{Shirai2014} in a 2x2x2 supercell with the same recipe (14.05 at.\% C) for comparison. The model consists of 3 units of B$_{11}$C$^{p}$(CBC), a unit of B$_{12}$(B$_{4}$), and 4 units of B$_{12}$(CBC). The total energy differences of the SA-SQS model II, relative to B$_{12}$(CBC) and a chemical potential of diamond, are given in TABLE~\ref{tab:4}. Compared to the results in the literature, our results show a fairly good agreement. In fact, the total energies of the SA-SQS model II are somewhat lower due to the removal of cell constraints. However, having the icosahedral carbon atoms at the equatorial sites, bonded to the 4-atom rhombic chains (Model II/(\emph{e})), in our case does not reduce the total energy as reported in the literature~\cite{Shirai2014}. We underline that the negative values of $\Delta$E in TABLE~\ref{tab:4} are no proof of stability in thermodynamic sense. To examine their thermodynamic stability, one might need to consider also the energy difference with respect to B$_{4}$C. The electronic DOS for the SA-SQS model II are illustrated by Fig.~\ref{fig:9}(a) and \ref{fig:9}(b). Both of them have the same metallic character and the same hole density of 0.25 hole/s.a. as the model II in~\cite{Shirai2014}. Even though the distances between the valence band edge and the first gap state are different from that in the literature, the qualitatively corresponding results for the model II are achieved with the SA-SQS approach, strengthening the reliability of the approach.}\\
\begin{table}
\caption{Energy difference ($\Delta$E) of Model II proposed in Ref.~\cite{Shirai2014}, relative to B$_{12}$(CBC) plus a chemical potential of diamond, as a correction for one additional carbon atom. As given in Ref.~\cite{Shirai2014}, type (\emph{p}) indicates configurations, in which all icosahedral carbon atoms are located at the polar site, meanwhile (\emph{e}) indicates configurations, in which at least one of the icosahedral carbon atoms is located at the equatorial site.}
\begin{tabular}{l @{\hspace{0.5cm}} c @{\hspace{0.5cm}} c}
\hline
\hline
Configuration/(type) & $\Delta$E (eV/atom)\\
\hline
Model II/(\emph{p}) & -0.004 \footnote{This work (GGA-PAW without cell constraint)\label{note:a}} \\
 & 0.007 (IId) \footnote{Ref.~\cite{Shirai2014} - Shirai \emph{at al.} (LDA-PP with cell constraints)\label{note:b}}  \\
 & 0.011 (IIe) \footref{note:b} \\
\hline
Model II/(\emph{e}) & -0.004 \footref{note:a}\\
 & 0.004 (IIf) \footref{note:b} \\
 & 0.001 (IIg) \footref{note:b} \\
\hline
\hline
\end{tabular}
\label{tab:4}
\end{table}
        
\indent{We then construct another SA-SQS model of B$_{13}$C$_{2}$, based the three types of superatom in Fig.~\ref{fig:8}, in a 4x4x3 supercell. The cell consists of 48 superatoms (735 atoms): 32 units of B$_{11}$C$^{p}$(CBC), 15 units of B$_{12}$(B$_{4}$), and a unit of B$_{12}$(CBC). This is also inspired by the suggestion of Shirai \emph{et al.}~\cite{Shirai2014}, but in contrast to that work, our larger supercell contains exact stoichometry of B$_{13}$C$_{2}$. As shown in Fig.~\ref{fig:9}(c), this SA-SQS model of B$_{13}$C$_{2}$, namely Rh-B$_{13}$C$_{2}$, is a semiconductor with a GGA-PBE96-based band-gap of 1.42 eV. The formation energy, calculated from $\alpha$-boron and diamond is -0.049 eV/atom. This is higher than that of (B$_{12}$(CBC))$_{0.5}$(B$_{11}$C$^{e}$(BBC))$_{0.5}$ of the \emph{e}(1-6) configuration by 0.027 eV/atom at 0 K. We note that to determine the configurational entropy, in this case, we take into account the possibility of having (1) the substitution of icosahedral carbon atoms at different polar sites and (2) three fold rotational degeneracy of the 4-atom rhombic chains to obtain an upper bound of the configurational entropy. Nevertheless, Rh-B$_{13}$C$_{2}$ is not favorable with respect to the \emph{e}(1-6) configuration at any realistic temperature below the extreme 17000 K. The total energy of Rh-B$_{13}$C$_{2}$ can be lowered, as shown in Ref.~\cite{Shirai2014}, by increasing a concentration of the B$_{12}$(CBC) ground state units. However, one can expect that by doing so the compound would be more metallic. Consequently, based on our results, we propose that at 13.33 at.\% C, the atomic configuration of boron carbide is dominated by a mixture of the B$_{12}$(CBC) and the B$_{11}$C$^{e}$(BBC) units, although other structural units, like B$_{11}$C$^{p}$(BBC), B$_{11}$C$^{p}$(CBC) and B$_{12}$(B$_{4}$) can be present at lower concentrations.}\\
\begin{table}
\caption{Formation energy ($\Delta$E$^{form}$) of different 15-atom B$_{14}$C units, with respect to $\alpha$-boron and diamond.}
\begin{tabular}{l @{\hspace{0.5cm}} c @{\hspace{0.5cm}} c}
\hline
\hline
Structural unit & $\Delta$E$^{form}$ (eV/atom)\\
\hline
B$_{12}$(BBC) & 0.010\\
B$_{11}$C$^{p}$(BBB) & 0.067\\
B$_{11}$C$^{e}$(BBB) & 0.078\\
B$_{12}$(BCB) & 0.211\\
\hline
\hline
\end{tabular}
\label{tab:5}
\end{table}
\section{VII. Implications on understanding other boron carbides}
\indent{Based on our model proposed in this work, as well as our previous study of B$_{4}$C~\cite{Annop2014}, we can suggest the following scenario for structural evolution as the stoichiometry is deviating from the two "ideal" B$_{13}$C$_{2}$ and B$_{4}$C compositions: going from B$_{13}$C$_{2}$ toward the carbon-rich limit ($\thicksim$20 at.\% C, or B$_{4}$C), the B$_{12}$(CBC) and the B$_{11}$C$^{e}$(BBC) units would gradually be replaced by the stable, semiconducting B$_{11}$C$^{p}$(CBC) units with configurational disorder of carbon atoms on the icosahedral polar sites~\cite{Annop2014}, until the latter dominates the boron carbide system at the carbon-rich limit as suggested in the literature~\cite{Bylander1990, Vast1999, Mauri2001, Annop2014}. On the other hand, if the carbon concentration becomes lower than 13.33 at.\% C toward the boron-rich limit ($\thicksim$8 at.\% C), an equal fraction of carbon atoms in the B$_{11}$C$^{e}$ icosahedra and the (CBC) chains could be substituted by boron atoms, thus yielding the B$_{12}$(BBC) units, which we find to have the lowest formation energy (0.01 eV/atom), among the other B$_{14}$C units (see TABLE~\ref{tab:5}), with respect to $\alpha$-boron and diamond. As shown in section IV, the B$_{12}$(BBC) unit is a semiconductor. Thus, based on this structural model, boron carbide would maintain semiconducting character within its whole single-phase region due to specific types of low energy configurational disorder stabilized by entropy as shown in detail for B$_{13}$C$_{2}$ in the present work. However, apart from the stoichiometries B$_{4}$C and B$_{13}$C$_{2}$, details of the stability and the properties of other boron carbide compositions deserve further investigation.}
\section{VIII. Conclusions}
\indent{We investigate the properties of B$_{13}$C$_{2}$ using first-principles calculations, in which substitutionally disordered configurations are modelled within the SA-SQS scheme. The calculations of dilute stoichiometric defects reveal, in the matrix of the presumed ground state B$_{13}$C$_{2}$ represented by the B$_{12}$(CBC) units, the B$_{11}$C$^{e}$(BBC) unit have the lowest formation energy. We thus predict that as the temperature increases, the low temperature-ordered B$_{13}$C$_{2}$, comprising only of the B$_{12}$(CBC) units, becomes thermodynamically unfavorable with respect to substitutional disorder, in which some of the B$_{12}$(CBC) units turn into the B$_{11}$C$^{e}$(BBC) units, leading to substitutionally disordered B$_{13}$C$_{2}$. The splitting of valence states in the B$_{11}$C$^{e}$(BBC) units then compensates partially or fully the electron deficiency in the B$_{12}$(CBC) units, depending on their relative concentrations. Consequently, as the concentration of the B$_{11}$C$^{e}$(BBC) units increases with temperature under equilibrium condition, approaching 50\%, the electron deficiency in B$_{13}$C$_{2}$ is gradually compensated until the completely semiconducting character is achieved. Also the calculated band-gap for disordered B$_{13}$C$_{2}$ is in fairly good agreement with the experiment. It is possible that the metallic ordered B$_{13}$C$_{2}$ has not been experimentally observed because of the limited atomic diffusion in boron carbide even at elevated temperature, thus freezing in the high temperature disordered configurations of B$_{13}$C$_{2}$ also at intermediate temperature.}\\
\indent{This mixture of the B$_{12}$(CBC) and the B$_{11}$C$^{e}$(BBC) units is found to have lower total energy and Gibbs free energy than a model based on previously suggested B$_{11}$C$^{p}$(CBC), B$_{12}$(B$_{4}$), and B$_{12}$(CBC) units~\cite{Shirai2014}. The model for structural disorder proposed in the present work can explain experimental finding of semiconducting character of B$_{13}$C$_{2}$.}
\section{IX. Acknowledgments}
Financial support by the Swedish Research Council (VR) through the young researcher grant No. 621-2011- 4417 and the international career grant No. 330-2014-6336 is gratefully acknowledged by B.A. The support from CeNano at Linköping University is acknowledged by A.E. and B.A.The support from Swedish Research Council (VR) Project No. 2014-4750, LiLi-NFM, and the Swedish Government Strategic Research Area Grant in Materials Science to the AFM research environment at LiU are acknowledged by S.I.S. The simulations were carried out using supercomputer resources provided by the Swedish national infrastructure for computing (SNIC) performed at the National supercomputer centre (NSC). Sit Kerdsongpanya is acknowledged for useful discussions. 


\begin{thebibliography}{10}

\bibitem{thevenot1990}
F.~Thev{\'e}not.
\newblock {Boron carbide - A comprehensive review}.
\newblock {\em Journal of the European Ceramic Society}, 6:205, 1990.

\bibitem{Emin2005}
D.~Emin and T.~L. Aselage.
\newblock {A proposed boron-carbide-based solid-state neutron detector}.
\newblock {\em Journal of Applied Physics}, 97:013529, 2005.

\bibitem{Caruso2006}
A.~N. Caruso, P.~A. Dowben, S.~Balkir, N.~Schemm, K.~Osberg, R.~W. Fairchild,
  O.~B. Flores, S.~Balaz, A.~D. Harken, B.~W. Robertson, and J.~I. Brand.
\newblock {The all boron carbide diode neutron detector: Comparison with
  theory}.
\newblock {\em Materials Science and Engineering B}, 135:129, 2006.

\bibitem{Lacy2011}
J.~L. Lacy, A.~Athanasiades, L.~Sun, C.~S. Martin, T.~D. Lyons, M.~A. Foss, and
  H.~B. Haygood.
\newblock {Boron-coated straws as a replacement for $^{3}$He-based neutron
  detectors}.
\newblock {\em Nuclear Instruments and Methods in Physics Research A}, 652:359,
  2011.

\bibitem{Carina2012}
C.~H{\"o}glund, J.~Birch, K.~Andersen, T.~Bigault, J.-C. Buffet, J.~Correa,
  P.~van Esch, B.~Guerard, R.~Hall-Wilton, J.~Jensen, A.~Khaplanov,
  F.~Piscitelli, C.~Vettier, W.~Vollenberg, and L.~Hultman.
\newblock {B$_{4}$C thin films for neutron detection}.
\newblock {\em Journal of Applied Physics}, 111:104908, 2012.

\bibitem{Clark1943}
H.~K. Clark and J.~L. Hoard.
\newblock {The Crystal Structure of Boron Carbide}.
\newblock {\em Journal of the American Chemical Society}, 65:2115, 1943.

\bibitem{Yakel1975}
H.~L. Yakel.
\newblock {The Crystal Structure of a Boron-Rich Boron Carbide}.
\newblock {\em Acta Crystallographica Section B}, 31:1797, 1975.

\bibitem{Morosin1986}
B.~Morosin, A.~W. Mullendore, D.~Emin, and G.~A. Slack.
\newblock {Rhombohedral crystal structure of compunds containing boron-rich
  icosahedra}.
\newblock {\em AIP Conference Proceedings}, 140:70, 1986.

\bibitem{Morosin1995}
B.~Morosin, G.~H. Kwei, A.~C. Lawson, T.~L. Aselage, and D.~Emin.
\newblock {Neutron powder diffraction refinement of boron carbides. Nature of
  intericosahedral chains}.
\newblock {\em Journal of Alloys and Compounds}, 226:121, 1995.

\bibitem{Gosset1991}
D.~Gosset and M.~Colin.
\newblock {Boron carbides of various compositions: An improved method for
  X-rays characterization}.
\newblock {\em Journal of Nuclear Materials}, 183:161, 1991.

\bibitem{Domnich2011}
V.~Domnich, S.~Reynaud, R.~A. Haber, and M.~Chhowalla.
\newblock {Boron Carbide: Structure, Properties, and Stability under Stress}.
\newblock {\em Journal of the American Ceramic Society}, 94(11):3605, 2011.

\bibitem{Bylander1990}
D.~M. Bylander, L.~Kleinman, and S.~Lee.
\newblock {Self-consistent calcualtions of the energy bands and bonding
  properties of B$_{12}$C$_{3}$}.
\newblock {\em Physical Review B}, 42(2):1394, 1990.

\bibitem{Vast1999}
R.~Lazzari, N.~Vast, J.~M. Besson, S.~Baroni, and A.~D. Corso.
\newblock {Atomic structure and vibrational properties of icosahedral B$_{4}$C
  boron carbide}.
\newblock {\em Physical Review Letters}, 83:3230, 1999.

\bibitem{Mauri2001}
F.~Mauri, N.~Vast, and C.~J. Pickard.
\newblock {Atomic Structure of Icosahedral B$_{4}$C Boron Carbide from a First
  Principles Analysis of NMR Spectra}.
\newblock {\em Physical Review Letters}, 87(8):085506, 2001.

\bibitem{Annop2014}
A.~Ektarawong, S.~I. Simak, L.~Hultman, J.~Birch, and B.~Alling.
\newblock {First-principles study of configurational disorder in B$_{4}$C using
  a superatom-special quasirandom structure method}.
\newblock {\em Physical Review B}, 90:024204, 2014.

\bibitem{Bylander1991}
D.~M. Bylander and L.~Kleinman.
\newblock {Structure of B$_{13}$C$_{2}$}.
\newblock {\em Physical Review B}, 43(2):1487, 1991.

\bibitem{Saal2007}
J.~E. Saal, S.~Shang, and Z.~K. Liu.
\newblock {The structural evolution of boron carbide via \emph{an initio}
  calculations}.
\newblock {\em Applied Physics Letters}, 91:231915, 2007.

\bibitem{Vast2009}
N.~Vast, J.~Sjakste, and E.~Betranhandy.
\newblock {Boron carbides from first principles}.
\newblock {\em Journal of Physics: Conference Series}, 176:012002, 2009.

\bibitem{Shirai2014}
K.~Shirai, K.~Sakuma, and N.~Uemura.
\newblock {Theoretical study of the structure of boron carbide
  B$_{13}$C$_{2}$}.
\newblock {\em Physical Review B}, 90:064109, 2014.

\bibitem{Aselage1992}
T.~L. Aselage and R.~G. Tissot.
\newblock {Lattice constants of boron carbides}.
\newblock {\em Journal of the American Ceramic Society}, 75(8):2207, 1992.

\bibitem{Tallant1989}
D.~R. Tallant, T.~L. Aselage, A.~N. Campbell, and D.~Emin.
\newblock {Boron carbide structure by Raman spectroscopy}.
\newblock {\em Physical Review B}, 40(8):5649, 1989.

\bibitem{Aselage1997}
T.~L. Aselage, D.~R. Tallant, and D.~Emin.
\newblock {Isotope dependencies of Raman spectra of B$_{12}$As$_{2}$,
  B$_{12}$P$_{2}$, B$_{12}$O$_{2}$, and B$_{12+x}$C$_{3-x}$: Bonding of
  intericosahedral chains}.
\newblock {\em Physical Review B}, 56(6):3122, 1997.

\bibitem{Aselage1998}
T.~L. Aselage and D.~R. Tallant.
\newblock {Association of broad icosahedral Raman bands with substitutional
  disorder in SiB$_{3}$ and boron carbide}.
\newblock {\em Physical Review B}, 57(5):2675, 1998.

\bibitem{Arm1983}
D.~R. Armstrong, J.~Bolland, P.~G. Perkins, G.~Will, and A.~Kirfel.
\newblock {The nature of the chemical bonding in boron carbide. IV. Electronic
  band structure of boron carbide, B$_{13}$C$_{2}$, and three model of the
  structure B$_{12}$C$_{3}$}.
\newblock {\em Acta Crystallographica Section B}, 39:324, 1983.

\bibitem{Vast2004}
M.~Calandra, N.~Vast, and F.~Mauri.
\newblock {Superconductivity from doping boron icosahedra}.
\newblock {\em Physical Review B}, 69:224505, 2004.

\bibitem{Wood1984}
C.~Wood and D.~Emin.
\newblock {Conduction mechanism in boron carbide}.
\newblock {\em Physical Review B}, 29:4582, 1984.

\bibitem{Zuppiroli1991}
L.~Zuppiroli, N.~Papandreou, and R.~Kormann.
\newblock {The dielectric response of boron carbide due to hopping conduction}.
\newblock {\em Journal of Applied Physics}, 70:246, 1991.

\bibitem{Musiri2007}
M.~M. Balakrishnarajan, P.~D. Pancharatna, and R.~Hoffmann.
\newblock {Structure and bonding in boron carbide: The invicibility of
  imperfections}.
\newblock {\em New Journal of Chemistry}, 31:473, 2007.

\bibitem{Werheit2000}
R.~Schmechel and H.~Werheit.
\newblock {Structural defects of some icosahedral boron-rich solids and their
  correlation with the electronic properties}.
\newblock {\em Journal of Solid State Chemistry}, 154:61, 2000.

\bibitem{Werheit2007}
H.~Werheit.
\newblock {Are there bipolarons in icosahedral boron-rich solids?}
\newblock {\em Journal of Physics: Condensed Matter}, 19:186207, 2007.

\bibitem{Zunger1990}
A.~Zunger, S.-H. Wei, L.~G. Ferreira, and J.~E. Bernard.
\newblock {Special Quasirandom Structures}.
\newblock {\em Physical Review Letters}, 65(3):353, 1990.

\bibitem{Kresse1996}
G.~Kresse and J.~Furthm{\"u}ller.
\newblock {Efficiency of ab-initio total energy calculations for metals and
  semiconductors using a plane-wave basis set}.
\newblock {\em Computational Material Science}, 6:15, 1996.

\bibitem{Kresse19962}
G.~Kresse and J.~Furthm{\"u}ller.
\newblock {Efficient iterative schemes for $\emph{ab initio}$ total-energy
  calculations using a plane-wave basis set}.
\newblock {\em Physical Review B}, 54:11169, 1996.

\bibitem{Blochl1994}
P.~E. Bl\"ochl.
\newblock {Projector augmented-wave method}.
\newblock {\em Physical Review B}, 50(24):17953, 1994.

\bibitem{Perdew1996}
J.~Perdew, K.~Burke, and M.~Ernzerhof.
\newblock {Generalized Gradient Approximation Made Simple}.
\newblock {\em Physical Review Letters}, 77:3865, 1996.

\bibitem{Becke2006}
A.~D. Becke and E.~R. Johnson.
\newblock {A simple effective potential for exchange}.
\newblock {\em Journal of Chemical Physics}, 124:221101, 2006.

\bibitem{Tran2009}
F.~Tran and P.~Blaha.
\newblock {Accurate Band Gaps of Semiconductors and Insulator with a Semilocal
  Exchange-Correlation Potential}.
\newblock {\em Physical Review Letters}, 102:226401, 2009.

\bibitem{HSE03}
J.~Paier, M.~Marsman, K.~Hummer, G.~Kresse, I.~C. Gerber, and J.~G.
  $\acute{A}$ngy$\acute{a}$n.
\newblock {Screened hybrid density functional applied to solids}.
\newblock {\em The Journal of Chemical Physics}, 124:154709, 2006.

\bibitem{HSE06}
A.~V. Krukau, O.~A. Vydrov, A.~F. Izmaylov, and G.~E. Scuseria.
\newblock {Influence of exchange screening parameter on the performance of
  screened hybrid functionals}.
\newblock {\em The Journal of Chemical Physics}, 125:224106, 2006.

\bibitem{Monkhorst1976}
H.~J. Monkhorst and J.~D. Pack.
\newblock {Special points for Brillouin-zone integrations}.
\newblock {\em Physical Review B}, 13:5188, 1976.

\bibitem{Vesta2011}
K.~Momma and F.~Izumi.
\newblock {$\emph{VESTA 3}$ for three-dimensional visualization of crystal,
  volumetric and morphology data}.
\newblock {\em Journal of Applied Crystallography}, 44:1272, 2011.

\bibitem{Kirfel1979}
A.~Kirfel, A.~Gupta, and G.~Will.
\newblock {The nature of the chemical bonding in boron carbide,
  B$_{13}$C$_{2}$. I. Structure Refinement}.
\newblock {\em Acta Crystallographica Section B}, 35:1052, 1979.

\bibitem{Werheit2006}
H.~Werheit.
\newblock {On excitons and other gap states in boron carbide}.
\newblock {\em Journal of Physics: Condensed Matter}, 18:10655, 2006.

\bibitem{Isaev2011}
E.~I. Isaev, S.~I. Simak, A.~S. Mikhaylushkin, Yu.~Kh. Vekilov, E.~Yu.
  Zarechnaya, L.~Dubrovinsky, N.~Dubrovinskaia, M.~Merlini, M.~Hanfland, and
  I.~A. Abrikosov.
\newblock {Impact of lattice vibrations on equation of state of the hardest
  boron phase}.
\newblock {\em Physical Review B}, 83:132106, 2011.

\bibitem{Yao2014}
S.~Yao, W.~P. Huhn, and M.~Widom.
\newblock {Phase transitions of boron carbide: Pair interaction model of high
  carbon limit}.
\newblock {\em Solid State Sciences}, 47:21, 2015.

\bibitem{Morosin1996}
G.~H. Kwei and B.~Morosin.
\newblock {Structure of the Boron-Rich Boron Carbides from Neutron Powder
  Diffraction: Implications for the Nature of the Inter-Icosahedral Chains}.
\newblock {\em Journal of Physical Chemistry}, 100:8031, 1996.

\bibitem{Larson1986}
A.~C. Larson.
\newblock {Comments Concerning the Crystal Structure of B$_{4}$C}.
\newblock {\em AIP Conference Proceedings}, 140:109, 1986.

\bibitem{Kuzen1979}
M.~A. Kuzenkova, P.~S. Kislyi, B.~L. Grabchuk, and N.~I. Bodnaruk.
\newblock {The structure and properties of sintered boron carbide}.
\newblock {\em Journal of the Less-Common Metals}, 67:217, 1979.

\bibitem{Hautier2012}
G.~Hautier, S.~P. Ong, A.~Jain, C.~J. Moore, and G.~Ceder.
\newblock {Accuracy of density functional theory in predicting formation
  energies of ternary oxides from binary oxides and its implication on phase
  stability}.
\newblock {\em Physical Review B}, 85:155208, 2012.

\bibitem{Alling2014}
B.~Alling.
\newblock {Metal to semiconductor transition and phase stability of
  Ti$_{1-x}$Mg$_{x}$N$_{y}$ alloys investigated by first-principles
  calculations}.
\newblock {\em Physical Review B}, 89:085112, 2014.

\bibitem{Alling2011}
B.~Alling, A.~V. Ruban, A.~Karimi, L.~Hultman, and I.~A. Abrikosov.
\newblock {Unified cluster expansion method applied to the configurational
  thermodynamics of cubic Ti$_{1-x}$Al$_{x}$N}.
\newblock {\em Physical Review B}, 83:104203, 2011.

\bibitem{Garbulsky1996}
G.~D. Garbulsky and G.~Ceder.
\newblock {Contribution of the vibrational free energy to phase stability in
  substitutional alloys: Methods and trends}.
\newblock {\em Physical Review B}, 53(14):8993, 1996.

\end{thebibliography}

\end{document}